# On the Use of AI for Satellite Communications

Miguel Ángel Vázquez[1], Pol Henarejos[1], Ana I. Pérez-Neira[1,2], Elena Grechi[3], Andreas Voight[3], Juan Carlos Gil[4], Irene Pappalardo[5], Federico Di Credico[5], Rocco Michele Lancellotti[5]

Centre Tecnològic de Telecomunicacions de Catalunya, Castelldefels (Spain) [1].

Universitat Politècnica de Catalunya, Barcelona (Spain) [2].

Eutelsat, Paris (France) [3].

GMV Aerospace and Defense, Tres Cantos (Spain) [4].

Reply, Turin (Italy) [5].

*Abstract* —This document presents an initial approach to the investigation and development of artificial intelligence (AI) mechanisms in satellite communication (SatCom) systems. We first introduce the nowadays SatCom operations which are strongly dependent on the human intervention. Along with those use cases, we present an initial way of automatizing some of those tasks and we show the key AI tools capable of dealing with those challenges. Finally, the long term AI developments in the SatCom sector is discussed.

## INTRODUCTION

It may come as a surprise to realize that nowadays satellite communications still heavily depend on human expertise and manual operations. Satellite operator's teleports require strong human involvement, leading to a high operational expenditures (OPEX) and a reduced client quality of service. Indeed, ticketing processing of incidents in the radiofrequency (RF) plane require the intervention of human experts in order to provide a technical solution.

Moreover, with the forthcoming deployment of flexible payloads, satellites are meant to reconfigure their transmission budgets on a millisecond basis. This requires radio resource management algorithms with a very short response time, able to design time, frequency, power and beamforming in order to meet the clients service level agreements. Furthermore, the advent of satellite megaconstellations only add complexity to the picture, imposing the need to coordinate space networks of hundreds if not thousands of satellites.

In this context, it is forecasted that automation algorithms will be incorporated to the daily satellite operations in the next few years. The use of deep learning may play an important role in this problem due to its capability of mimicking any nonlinear function.

In this work we introduce four use cases; namely anomaly detection in telemetry data, flexible payload optimization, interference detection and classification and beam congestion prediction, which we consider that AI may have a substantial impact on. All of them are part of nowadays and near future satellite communications daily operations. AI tools capable of dealing with the described use cases are introduced. In addition, other ideas on how AI can support long term satellite communications are also presented.

## PROMISING USE CASES

### A. Anomaly Detection in Telemetry data

Conventional telemetry usage very often refers to comparison between the instantaneous readings and static thresholds. Given the large amount of data downlinked from the satellites, it is often difficult to analyse trends or detect anomalies if readings do not exceed these thresholds. The introduction of AI and a new approach including data labelling would bring considerable benefits to the industry to better identify anomalies including not only failures but patterns deviating from nominal behaviours. By labelling data referring to a specific event as an anomaly, specific future failures could be prevented in a timely manner potentially avoiding heavy on board reconfigurations, but acting in a more proactive approach. At industry level, anomaly early detection could allow, for instance, spare on board equipment saving such as traveling-wave tube amplifiers (TWTAs) or could even trigger a better ground service customer management and TWTA gains configuration according to telemetry trends.

This use case can be described as the generic ML problem of anomaly detection in multivariate temporal data series. Despite it being a topic which has received considerable attention in the last 10 years [1], its applicability to real systems is subject to investigation in current innovation labs within companies and academic groups.

There is a myriad of options for addressing the problem of anomaly detection in multidimensional time series. One case is based on robust principal component analysis by Netflix [2] on payment transactions. Yet another case is the usage of predictive statistical methods for detecting DDoS attacks as reported in [3].

### B. *Flexible Payload Optimization*

Flexible payloads will become mainstream in the near future and will revolutionize the conventional idea of a satellite communications mission [4]. On top of the new perspective of reconfigurable payload, AI shall be integrated in the end-to-end service delivered in order to achieve a smart interference management system. A fully reconfigurable payload operated in a conventional approach with regards to interference management could not bring any substantial differentiator to the industry neither to customers, who on the other hand are expected to pay for more expensive capacity. Instead, the winning combination of flexible payload with smart interference avoidance system is definitely a must that can not only deliver the benefits of the previously mentioned use cases but also increase customer satisfaction and stickiness.

The main idea of this use case is to develop a mechanism able to provide a satellite payload configuration in terms of power allocation per feed element, bandwidth allocation and beamforming design, given a sudden change in the interference power level of a certain coverage area. The new payload configuration shall be provided in a very short time lapse; thus, it follows that the mechanism shall manifest a reduced computational complexity.

Existing techniques use parallel processing or other high performance computation means to explore every possibility. During the deployment, the applications require significant engineering effort to configure the software to match specific payloads. For this use, we expect to revisit the recent activities of deep learning applications to operations research [5], where deep neural networks are used for either reducing the optimization search space or providing initial efficient configurations.

Using a learning system we would expect the following gains: i)Less engineering effort would be needed to define payload configuration, therefore less effort to set up each instance. ii)Faster computation with no need to assess every possible solution. As payloads get more complex this factor becomes more and more relevant and may enable automated reconfiguration. iii)Can include factors that need to be separately programmed into a mass computation system (e.g. routing restrictions or "do not touch" components)

### C. *Interference Detection and Classification*

Unfortunately, interferences in satellite communications are usually frequent. Interference detection is typically a task performed on a reactive mode instead of proactive. Given that in most of the cases satellites simply rely signals coming from the Earth, interfering signals are present to a large extent in all frequency bands. The possibility to offload a purely human task such as Power Spectrum Density check to an automated system, able to detect the presence of unwanted signals, is an exciting perspective in terms of improved spectral management and customer incident avoidance.

Most of the interferences present today are caused by human errors – either due to mispointed antenna (cross-polarization or adjacent satellite) or misconfigured equipment (noise introduction, intermodulation, etc.). These parameters cover 70-80% of all interference cases and are not related terrestrial networks. If there are carrier overlaps, it implies a digital video subscriber (DVB) carrier overlapping another DVB carrier or a satellite modem transmission, including very small aperture terminal (VSAT) traffic in time division multiple access (TDMA), for instance.

In order to mitigate them, there are several techniques that may help to reduce the levels of interferences. However, in many occasions, it is still difficult to cope with them. Currently, the only way to manage interference is by human intervention and performing an exhaustive analysis, that may take several days to solve the incidence. In other words, there are qualified personnel dedicated to detect interferences by inspecting figures, such as the spectrum, abnormal error rates increase or degraded user experience.

Here we propose an automated system capable to detect interferences and reduce the human intervention in order to generate alarms when interferences are occurring to perform operations in an automated fashion. The proposed system implements latest improvements on AI algorithms that are able to detect deviations of the signal's statistics, altered by unwanted interferences [6]. At the end, the system throws an alarm if the deviation trespasses a fixed threshold.

### D. *Congestion Prediction*

The load prediction problem is a widely analyzed topic. However, to the best of our knowledge, the application of machine learning algorithms to improve congestion prediction has yet to appear because of the difficult challenge in characterizing the satellite communication signals. This is especially useful for detecting anomalous behaviors and predicting not recurrent patterns in a strong non-linear scenario.

The most common approach to the satellite system congestion consists in two phases: performance prediction and application of mitigation techniques.

As of today, the performance prediction is in general based on trend curves of the previous days. This does not allow neither long term predictions nor management of anomalous behaviors. By considering several additional factors (e.g. traffic type in terms of services and download/upload bandwith), machine learning techniques can both extend the time frame on which the analysis is done and identify in quasi-real time irregular traffic patterns that unexpectedly affect the performance (e.g. release of os update or live events). A long time frame prediction is particularly useful for marketing purposes like identifying seasonal trends and simulating future scenarios characterized by a different number of users.

On the other hand, mitigation techniques are commonly based on human decisions and therefore they show scalability issues as well as a limited view on the effect of the applied actions. Also, in this phase, machine learning techniques can boost the effectiveness of the countermeasures, implementing automatic and real-time mechanisms to cope with congestion impairments.

Note that meaningful predictions can only be made considering homogenous populations, which can be obtained through clustering algorithms. These are unsupervised algorithms that should pre-process the input data to the learning network in order to recognize hidden patterns in the data. Different existing techniques, hierarchical clustering, K-Means, Density-Based Spatial Clustering of Applications with Noise (DBSCAN), affinity propagation, and spectral clustering (especially suitable when the hidden structure does not obey to convex sets) are explored and results compared. Moreover, within arising clusters, forecasting and regression techniques (such as SARIMA, LSTM and XGBoost [4]) are used to predict future traffic levels.

## OTHER PROMISING USE CASES

5G satellite-terrestrial interoperability and IoT (Internet of Things) services can also benefit from AI techniques that can cope with big volumes of very diverse data (e.g. telemetry, communication, weather, earth observation, …), which present uncertainty due to data inconsistency and incompleteness, ambiguities, latency, model approximations, structured, unstructured text, multimedia,... In general, AI is becoming one of the main enablers for new services and to optimize industrial processes, reduce cost of available telecommunication technologies and systems, still pursuing performance increase in data rate and flexibility.

Last but not least, mega-constellation systems offering capacity to mobile users such as vessels and airplanes have to tackle a two-level dynamic environment: the satellite and the user terminals. This varying system has to be able to deal with potential hotspots beams (i.e. a coverage area which a high demand) that change over time and satellite. Although a classical data fusion between satellite and user terminals trajectories may help in detecting these hotspots, demands and motion present a stochastic nature which could be tackled via data-driven algorithms. In other words, the past detection of hotspots may support the inference of future ones.

## ACKNOWLEDGEMENT

*This work is funded by the European Space Agency through the ARTES Future Preparation programme. The view expressed herein can in no way be taken to reflect the official opinion of the European Space Agency.*